\documentclass[aps,prb,twocolumn,superscriptaddress]{revtex4}

\usepackage{graphicx}
\usepackage{dcolumn}
\usepackage{bm}
\usepackage{hyperref}

\begin{document}

\title{Localized basis sets for unbound electrons in nanoelectronics}

\author{D. Soriano}
\email{dsh2@alu.ua.es}
\affiliation{Departamento de F\'isica Aplicada, Universidad de Alicante, San Vicente del Raspeig, Alicante 03690, Spain}
\author{D. Jacob}
\email{djacob@physics.rutgers.edu}
\affiliation{Departamento de F\'isica Aplicada, Universidad de Alicante, San Vicente del Raspeig, Alicante 03690, Spain}
\affiliation{Department of Physics and Astronomy, Rutgers University, Piscataway NJ 08904 USA}
\author{J. J. Palacios}
\email{jj.palacios@ua.es}
\affiliation{Departamento de F\'isica Aplicada, Universidad de Alicante, San Vicente del Raspeig, Alicante 03690, Spain}

\date{\today}

\begin{abstract}
It is shown how unbound electron wave functions can be expanded in a suitably chosen localized basis sets for
any desired range of energies. In particular, we focus on the use of gaussian basis sets, commonly used in first-principles
codes. The possible usefulness of these basis sets in a first-principles description of field emission or scanning
tunneling microscopy at large bias is illustrated by studying a simpler related phenomenon: 
The lifetime of an electron in a H atom subjected to a strong electric field.
\end{abstract}

\pacs{71.15.Ap}

\maketitle

\section{INTRODUCTION}
The theoretical description
of the field emission (FE) mechanism has been traditionally based on the Wentzel-Kramers-Brillouin (WKB) 
approximation and simple band-structure models for the emitting tips\cite{Fowler-Nordheim}. 
Research on field-emitting graphitic compounds has
attracted much attention ever since its first observation
 in carbon nanotubes (CNT's) by de Heer et al.\cite{SCIENCE-268-845}. Because of their high aspect ratio as well
as mechanical and chemical stability, carbon nanotubes (CNT's) are regarded as
potential materials for field emitters and this has been experimentally demonstrated
in recent years\cite{bonard:918,coll:prb,PhysRevLett.89.197602}. 
The electronic structure of CNT's and other graphitic materials strongly depends on details at the atomic level and
not only on the overall shape. Therefore, 
an atomistic description of the emission process in these materials 
is pertinent. In this regard, tight-binding\cite{liang:245301}, static
density functional theory (DFT)\cite{PhysRevB.61.9986,escn:kim,fcn:kim}, 
or time-dependent density functional theory (TDDFT)\cite{fegr:wat,PhysRevB.65.085405}
calculations have been able to explore 
the theoretical possibilities of some graphitic compounds as field emitters.

In the last decade a great deal of work has also been devoted to master  electronic transport at the nanoscale. 
From a computational viewpoint the development of  \textit{ab initio} codes for quantum transport calculations 
has revolutionized the field, since, for the first time,  reliable predictions 
are possible for the resistance of nanoscale objects when a current is driven 
through\cite{PhysRevLett.84.979,Palacios:prb:01,Palacios:prb:02,Brandbyge:prb:02,PhysRevB.63.121104}. 
Many of these codes are based on the  non-equilibrium Green's function formalism (NEGF) which, in its most common
implementation, requires the 
use of localized basis sets. This is an important drawback which hampers the straightforward applicability of these codes 
to describe FE processes where electron wave functions are unbound and extend into the vacuum.

A precise description of extended wave functions is also important when the electron 
transport takes place between electrodes or metallic tips that are significantly
 separated from each other. This is typically the case in scanning tunneling microscopy 
experiments where the metallic tip is several \r{A} away from the surface and the electron current 
becomes a tunneling current. When one goes from the contact regime to the tunneling regime 
the vacuum region becomes so wide that diffuse functions centered on the tip and the surface atoms cannot 
describe this region properly. Care must be taken complementing the basis set away from the tip and 
substrate\cite{blanc:prb} for typical tip-surface distances. This is particularly important when a large voltage is 
applied between tip and sample which can affect the shape of the tunneling barrier to the point that
electrons are emitted into the vacuum before reaching the substrate\cite{bee:prl}.

In this report we examine the possibilities and limitations behind the use of localized atomic basis sets for the 
numerical description of free or unbound electrons. 
We are interested in the further implementation of these basis sets within our first-principles quantum transport package
ALACANT (Alicante Ab initio Computation Applied to 
Nanotransport)\cite{Palacios:prb:01,Palacios:prb:02,Jacob:prb:06,Jacob:thesis:07} 
which interfaces GAUSSIAN03\cite{Gaussian:03} or CRYSTAL\cite{Crystal:03} packages.
This is why we focus on the use of non-orthogonal gaussian-type functions. 
In Sec. II we present a detailed analysis of the completeness of gaussian basis sets in one dimension 
within a desired range of energies.  In Sec. III we generalize the analysis to three-dimensional gaussian wave functions. 
In Sec. IV, as a simple illustration, we compute the 
lifetime of an electron in the 2$s$ state of an H atom in the presence of an electric field using an
appropriate basis set and compare with analytical calculations. Section V presents the conclusions.

\section{Basic theoretical considerations in one dimension}

We consider first a one-dimensional free electron system described by the
kinetic energy Hamiltonian $\hat{T}_x=\frac{1}{2}\frac{d^2}{dx^2}$. Here and 
in the following we use atomic units (A.U.) where the electron mass is set to one.
The eigenstates are thus propagating waves which in the coordinate representation
are given by $\psi_k(x)\propto\exp[ikx]$ with a real wavevector 
$k$ and an energy dispersion given by
\begin{equation}
  \label{E-k-exact}
  E(k)=\frac{k^2}{2}.
\end{equation}
We want to describe free propagating waves with a basis set of localized atomic orbitals. 
As mentioned in the introduction, we focus on the case of gaussian-type functions which 
are employed by many first-principles
codes. In one dimension (1D) a gaussian-type function is given by:
\begin{equation}
  \label{1D-Gaussian}
  \phi(x)=\sqrt{2\alpha/\pi}\exp[-\alpha\,x^2].
\end{equation}
We now place an infinite number of gaussian orbitals on a regular one-dimensional grid with
lattice spacing $a$, and define the ket vector $|\phi_{n}\rangle$ corresponding to the 
gaussian wavefunction localized at grid point $x_n=na\,\vec{e}_x$ by $\langle x|\phi_{n}\rangle = \phi(x-na)$.
The overlap integral between two gaussian orbitals on the grid separated by a distance 
$na$ is thus given by
\begin{equation}
  \label{Overlap}
  s(na) := \langle \phi | \phi_n \rangle = \exp[ -\alpha (na)^2/2 ],
\end{equation}
For the kinetic energy integral between gaussian orbitals separated by a distance 
$na$ we have
\begin{eqnarray}
  \label{Hopping}
  t(na) &:=& \langle \phi | \hat{T}_x | \phi_n \rangle = -\frac{1}{2} \langle\phi | \hat{p}^2 | \phi_n \rangle
  \nonumber \\
  &=& \frac{\alpha}{2}(1-\alpha (na)^2)\exp[-\alpha (na)^2/2].
\end{eqnarray}
In this discrete basis set, the kinetic energy Hamiltonian $\hat{T}_x$ can be diagonalized
by transforming to Bloch states\cite{Ashcroft:book:76}:
\begin{equation}
  |\varphi_k\rangle =  \sum_{n=-\infty}^{\infty} \exp[ikna]\, |\phi_n\rangle 
  \Rightarrow \hat{T}_x|\varphi_k\rangle = \epsilon(k)|\varphi_k\rangle
\end{equation}
Multiplying the eigenvalue equation by $\langle\phi|$ from the left and 
resolving for $\epsilon(k)$, the following dispersion relation
is obtained:
\begin{equation}
  \label{E-k-approx}
  \epsilon(k)= \frac{\langle\phi|\hat{T}_x|\varphi_k\rangle}{\langle\phi|\varphi_k\rangle}
  = \frac{\sum_{n=-\infty}^{\infty} \exp[ikna] \, t(na)}{\sum_{n=-\infty}^{\infty}\exp[ikna]\, s(na)}.
\end{equation}

Obviously, the finer the grid, i.e. the smaller $a$, the better the dispersion relation must approach
that of free electrons.  In fact, in the limit of $a\rightarrow 0$, the expression (\ref{E-k-approx}) 
can be summed to yield the exact dispersion relation $E(k)$ {\it independent} of the gaussian 
exponent $\alpha$ (see Appendix):
\begin{equation}
  \label{limit}
  \lim_{a\rightarrow 0}\epsilon(k) = \frac{\int dx \, t(x) \exp[ikx] }{\int dx \, s(x) \exp[ikx]}
  = \frac{k^2}{2} \equiv E(k).
\end{equation} 
Of course, this is somewhat trivial in reality since on an infinitely fine grid any 
orbital must expand the entire Hilbert space. A text-book choice in this limit is a basis of 
delta functions (Dirac basis), which is obtained from the gaussian functions by taking $\alpha\rightarrow\infty$: 
\begin{equation}
  \lim_{\alpha\to\infty} \sqrt{\frac{2\alpha}{\pi}}\exp[-\alpha(x-na)^2] = \delta(x-na).
\end{equation}
\begin{figure}
  \includegraphics[width=\linewidth]{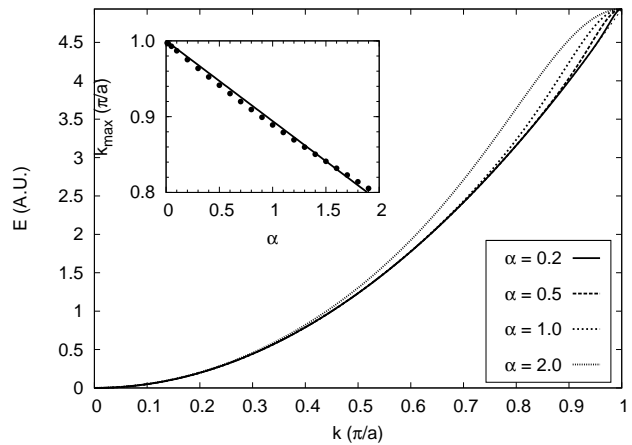}
  \caption{
    Energy dispersions $\epsilon(k)$ for $a=1$A.U. and for different values of $\alpha$.
    The inset shows the maximal k-value $k_{\rm max}$ calculated from the inflection point
    of the energy dispersion $\epsilon(k)$ as explained in the text. The points have been
    calculated from individual dispersion curves while the line is a linear fit 
    $k_{\rm max}(\alpha)=\pi/a-c\cdot\alpha$ to these points; $c\approx0.33 A.U.$.
  }
  \label{fig:dispersion}
\end{figure}

\begin{figure}
  \includegraphics[width=\linewidth]{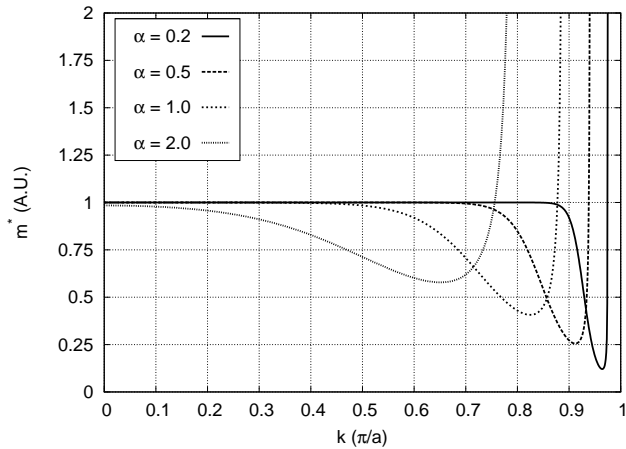}
  \caption{Effective mass $m^\ast(k)=(d^2\epsilon/dk^2)^{-1}$ for $a=1$A.U. and for different values of $\alpha$.}
  \label{fig:mass}
\end{figure}

However, for all practical purposes a finite grid parameter $a>0$ must be set. Our aim 
now is to find an optimal $\alpha$ for a given $a$. Note, that the ''atomistic'' 
description always yields a finite maximal $k$-vector --the Brillouin zone (BZ) boundary 
$k_{\rm BZ}=\pi/a$-- in contrast to truly free electrons where $k$ is unbound. Fig. 
\ref{fig:dispersion} shows the calculated dispersion $\epsilon(k)$ for fixed lattice 
parameter ($a=1$ A.U.) and for different values of $\alpha$. We see that in 
general the approximation works very well for smaller $k$-values, i.e. near the BZ center, 
but becomes worse towards the BZ boundary where the dispersion relation flattens out. 
Furthermore, the approximation is better, i.e. valid for a bigger range of $k$'s, the smaller 
$\alpha$, i.e. the more diffuse the Gaussian function. However, it is already remarkably 
good for relatively large values of $\alpha$. For instance, for $\alpha=1$, $\epsilon(k)$ is a good approximation 
for $k$-values up to $~70$\% of the theoretical upper limit $k_{\rm BZ}$. This can also be seen 
from the effective mass $m^\ast(k)=(d^2\epsilon/dk^2)^{-1}$ calculated numerically from the 
dispersion relation $\epsilon(k)$, and shown in Fig. \ref{fig:mass}. The smaller $\alpha$, the 
better does the effective mass $m^\ast$ approximate the constant mass of truely free electrons ($m_e=1$). 

This suggests that $\epsilon(k)\rightarrow E(k)$ for $\alpha\rightarrow 0$ for $k\le k_{\rm BZ}$, 
so that the optimal exponent $\alpha$ would be zero. Of course, setting $\alpha=0$ is numerically 
not feasible. In fact, in most quantum chemistry codes the gaussian exponents $\alpha$ cannot be 
chosen arbitrarily small for computational reasons. On the other hand, as can be seen from Fig. 
\ref{fig:dispersion}, the approximation is already quite good for reasonable values of $\alpha$ 
if we limit the range of wavevectors $k$ to some $k_{\rm max}$ smaller than the theoretical upper 
limit $k_{\rm BZ}$. Since the slope of the real free-electron dispersion increases linearly with 
$k$, $dE/dk=k$, we define $k_{\rm max}$ as the $k$-vector where the slope of the approximate energy 
dispersion $\epsilon(k)$ starts to decrease, i.e. at the inflection point of $\epsilon(k)$:
$d^2\epsilon/dk^2(k_{\rm max})\stackrel{!}{=}0$. Clearly, at this point the effective mass 
$m^\ast=(d^2\epsilon/dk^2)^{-1}$ becomes infinite, so that our approximation is not valid any longer. 
\cite{mass:comment}
As can be seen from the inset in Fig. \ref{fig:dispersion}, the thus calculated $k_{\rm max}$ can 
be fitted very well by a linear function of the gaussian exponent $\alpha$ that approaches the BZ 
boundary $k_{\rm BZ}=\pi/a$ for $\alpha\rightarrow0$.

How does $k_{\rm max}$ depend on the lattice parameter $a$? Suppose that for a given lattice parameter 
$a_0$ and exponent $\alpha_0$ we obtain a maximal range $k_{\rm max}^0$. Thus we have:
\[
\epsilon_{a_0,\alpha_0}(k_0) \approx  \frac{k_0^2}{2} \mbox{ for all } k_0 \le k_{\rm max}^0.
\] 
If we now alter the lattice spacing $a$ and simulaneously scale $\alpha$ as $\alpha(a)=\alpha_0(a_0/a)^2$ 
and $k$ as $k(a)=k_0(a_0/a)$ it is straightforward to show that  the energy dispersion is essentially 
unchanged apart from an overall factor of $(a_0/a)^2$:
\[
\epsilon_{a,\alpha}(k) = \left(\frac{a_0}{a}\right)^2 \epsilon_{a_0,\alpha_0}(k_0) 
\approx \left(\frac{a_0}{a}\right)^2 \frac{k_0^2}{2} = \frac{k^2}{2}
\]
Thus $\epsilon_{a,\alpha}(k) = k^2/2$ for all $k$ with $(a/a_0)k=k_0 \le  k_{\rm max}^0$,
i.e. for all $k$ with $k \le (a_0/a) k_{\rm max}^0\equiv k_{\rm max}$. Therefore $k_{\rm max}$ 
scales exactly as the theoretical limit $k_{\rm BZ}=\pi/a$ like $1/a$ with the lattice parameter 
$a$ if we simultaneosly scale the exponent $\alpha$ as $1/a^2$: $k_{\rm max}(a)=(a_0/a) k_{\rm max}^0$.

\section{Three-dimensional gaussian functions}

In three dimensions a gaussian wavefunction is given by
\begin{eqnarray}
  \label{eq:3D-Gaussian}
  \Phi(\vec{r}) &=& (2\alpha/\pi)^{3/2} \exp[-\alpha\,{\vec{r}}^2] 
  = \phi(x) \phi(y) \phi(z).
\end{eqnarray}
Again, we place the three-dimensional (3D) gaussian functions on a regular 1D grid along the x-axis with lattice
spacing $a$, and define the ket vector $|\Phi_n\rangle$ corresponding to the gaussian 
function localized at grid point $\vec{r}_n = n a \vec{e}_x$ by:
\begin{equation}
  \langle \vec{r} | \Phi_n \rangle = \Phi(\vec{r}-\vec{r}_n) = \phi(x-na) \phi(y) \phi(z)
\end{equation}
Now there is an additional contribution to the kinetic energy integral from the two directions 
($y$ and $z$) perpendicular to the direction ($x$) of the 1D grid:
\begin{eqnarray}
 \label{eq:kinetic-energy-3D}
  t_{\rm 3D}(na) &:=& \langle \Phi | (\hat{T}_x+\hat{T}_y+\hat{T}_z) | \Phi_{n} \rangle
  \nonumber\\
  &=& \langle\phi|\hat{T}|\phi_{na}\rangle \langle\phi|\phi\rangle^2
  + 2\times \langle\phi|\phi_{na}\rangle \langle\phi|\phi\rangle \langle\phi|\hat{T}|\phi\rangle 
  \nonumber\\
  &=& t(na) + 2 \times s(na) \times t(0) 
  \nonumber\\
  &=& t(na) + \alpha \, s(na).
\end{eqnarray}
Then, using  Bloch's theorem we obtain the corresponding dispersion relation for 3D gaussian functions
which differs from the analogous expression for 1D gaussian functions (\ref{E-k-approx}) only by a constant
energy shift:
\begin{equation}
  \label{E-k-approx-3D}
  \epsilon_{\rm 3D}(k) = \alpha + \frac{\sum_{n=-\infty}^{\infty} t(na) \exp[ikna]}
  {\sum_{n=-\infty}^{\infty} s(na)\exp[ikna]}.
\end{equation} 

The constant energy shift $\alpha$ is due to the lateral confinement of the electrons
to the region defined by the gaussian-shaped wavefunctions. Obviously, this confinement 
is eliminated by letting $\alpha\rightarrow0$, i.e. by increasing the diffuseness of 
the gaussian orbitals and thus letting the lateral wavefunction $\phi(y)\times\phi(z)$ 
become a free electron wave with wavevector zero ($k_y=k_z=0$). However, as said before, 
in most quantum chemistry packages the gaussian exponents cannot become arbitrarily small
for computational reasons. Thus when the artificial offset needs to be reduced beyond the 
computational limit for the exponents, one has to reduce the lateral confinement of the 
electrons by extending the vacuum basis set with gaussian wavefunctions along the other two 
dimensions. 

\begin{figure}
  \includegraphics[width=\linewidth]{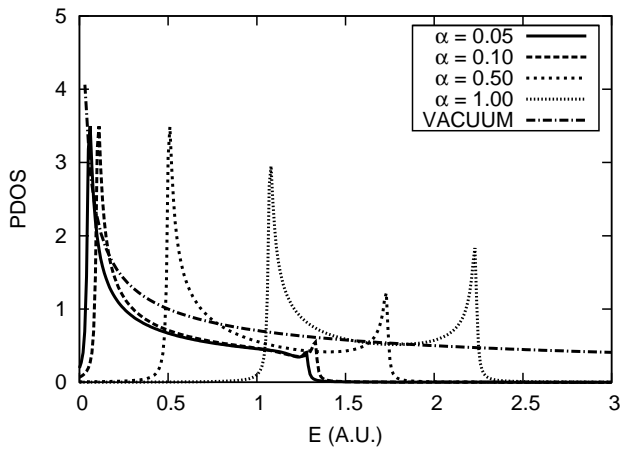}
  \caption{DOS projected onto a bulk site $i$ (PDOS) of a finite (N=2001) chain of gaussian functions for 
    different values of the gaussian exponent, $\alpha$, together with the DOS for free electrons 
    moving in one dimension.}
  \label{fig6}
\end{figure}

To analyze the density of states (DOS), we define the retarded Green's function in our 
non-orthogonal basis set as (for details see e.g refs.~\onlinecite{Economou:book:83,Jacob:thesis:07})
\begin{equation}
\widetilde{G}(E)=[(E+i\delta)S-T]^{-1},
\end{equation}
where $\delta$ is an infinitesimal quantity, 
$S$ is the overlap matrix, and $T$ is the 3D kinetic energy Hamiltonian represented 
in the non-orthogonal gaussian basis set.  
The integral over the total DOS, $\rho(E)$, must be equal to the number of basis functions, $N$, in our system.
\begin{equation}
  N = \int_{-\infty}^{\infty} dE\,\rho(E) =-\frac{1}{\pi}{\rm Im}\,\int_{-\infty}^{\infty} 
  dE\,{\rm Tr}[\widetilde{G}(E)\,S]
\end{equation}

In Fig. \ref{fig6} we plot the DOS of a finite chain of $N=2001$ gaussian functions
projected (PDOS) on a bulk site $i$, i.e. $-\frac{1}{\pi}{\rm Im}[\widetilde{G}(E)\,S]_{i,i}$, 
for four values of the gaussian exponent $\alpha$ and a fixed grid parameter of 
$a=2$ A.U.. The Green's function for an infinite system can be computed without difficulty\cite{Jacob:thesis:07}. We 
prefer, nevertheless, to present results for finite systems here since these are the ones employed in the next section.
Obviously, no differences should be expected for large enough systems.
As discussed above, there is an unwanted contribution to the kinetic energy coming from the confinement 
in the two directions perpendicular to the direction of the atomic chain. This opens a gap in the PDOS close to zero energy
which disappears as the exponent, and thus the lateral confinement, decreases.
For values of the exponent $0.05 \leq \alpha \leq 0.1$ the PDOS already                   
resembles that of free 1D electrons  for energy values from $0.0$ to $1.25$ A.U..   
The small but still visible gap in the PDOS at zero energy due to the residual lateral
confinement can be eliminated by letting $\alpha \to 0$, but, as already mentioned, numerical limitations inherent to  
most computational packages do not recommend to do this.
This already mentioned possibility of increasing the cross section of the chain 
by adding functions laterally will be explored in the future.

\section{IONIZATION OF THE 2$s$ STATE OF THE HYDROGEN ATOM}

\begin{figure}[tb]
  \includegraphics{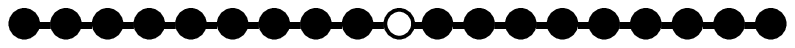}
  \caption{One dimensional model where a central 2s orbital of a hydrogen atom is connected to semi-infinite
    chains of gaussian functions on both sides representing the vacuum. 
    The electric field is applied along the chain direction.}
  \label{fig4}
\end{figure}

In order to test the use of gaussian basis sets in realistic FE calculations,  we have studied the ionization 
probability of the first excited state, $2s$, of the hydrogen atom in the presence of an electric field. 
For simplicity's sake, we have not included the $2p$ orbitals, although hybridization with them should be taken into account
in a more realistic description of this problem.
The calculations have been performed using 
the GAUSSIAN03 \emph{ab initio} package as a complementary test for situations when the use of 
standard {\em ab initio} packages is convenient or necessary.
The $2s$ orbital is coupled to two semi-infinite chains of gaussian functions representing the vacuum along the 
direction of the applied electric field as shown in Fig. \ref{fig4}. 
The straightforward use of GAUSSIAN03 forces us to work with finite vacuum chains as the one analyzed
in the previous section. The results presented below have been converged in the length of the lateral chains ($N=1000$)
so that no finite size effects can be appreciated. 

The lifetime of the $2s$ state in the presence of an electric field  can be extracted from the PDOS,
\begin{equation}
  \label{PDOS}
  \rho_{2s}(E)=-\frac{1}{\pi}{\rm Im}[\widetilde{G}(E)S]_{2s,2s}.
\end{equation}
The grid parameter and exponent for the two chains have been set to $1.0$ \r{A} ($\sim 2$ A.U.) and 
$0.07$ A.U., respectively. 
The distance between the hydrogen atom and both vacuum chains has been fixed to $5.0$ \r{A} ($\sim 10$ A.U.) 
to optimize the hybridization between the vacuum states and the $2s$ state of the hydrogen atom. 
The $2s$ atomic orbital is modeled by the corresponding STO-6G basis function implemented in GAUSSIAN03
 and the vacuum orbitals were implemented using ghost atoms. Care has been 
taken to remove the electron in the GAUSSIAN03 input file, reducing thus the calculation 
to a simple non-interacting problem. 

\begin{figure}
\includegraphics[width=\linewidth]{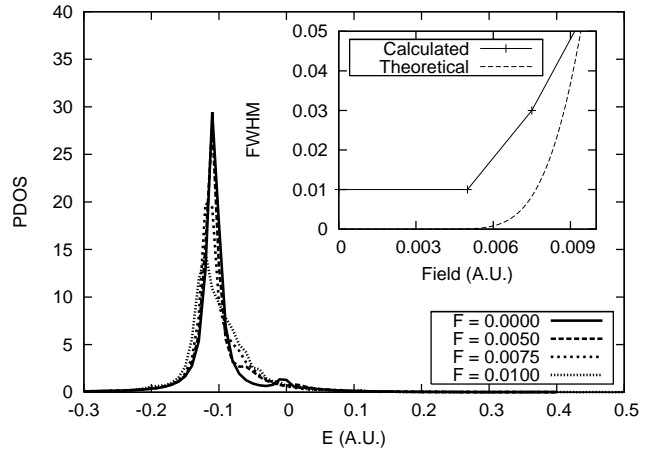}
\caption{Density of states projected on the $2s$ state of the hydrogen atom for different values of the
applied electric field. In the inset the field dependence of the peak width is depicted and 
compared with the theoretical one (dotted line).}
\label{fig7}
\end{figure}

The resonance width of the $2s$ orbital has been computed (see Fig.  \ref{fig7})
and compared to the theoretical one\cite{stark:see} for some values of the electric field F (see inset in Fig.\ref{fig7}).
A finite value of $\delta$ (or fictitious width) is included to smear out the computed PDOS. This is why our results for
the resonance width do not reach zero for zero applied field (see inset in Fig.\ref{fig7}).  
As it would be expected the computed resonance width $\Gamma_{2s}$ for the $2s$ state of the hydrogen atom is  well 
approximated by the quasi-classical theory:
\begin{equation}
\label{eqn10}
\Gamma_{2s}(F)=\frac{1}{32F^2}\textrm{ exp}\left(-\frac{1}{12F}\right)
\end{equation} 
Considering the simplifications in our calculation which has been restricted to one dimension and that little
care has been taken in completing the basis set around the $2s$ orbital, the results are qualitatively similar to
the theoretical ones. This illustrates that localized basis sets, when appropriately chosen, can be used to represent 
unbound electron wavefunctions which are required to describe FE phenomena or high-bias STM.

\section{conclusions}
In this work we have shown the possibility of representing
unbound electron states using localized gaussian-type
functions as a basis set. Although
plane waves are a more natural basis for unbound electrons,
localized basis sets have to be employed for the computation of field emission 
phenomena in the framework of many commonly used ab initio quantum transport packages such as ALACANT. 
In addition these basis sets  provide a more natural way of quantifying local atomic properties when needed.
As an illustration we have studied numerically the lifetime of an electron in the 2s orbital of the H atom 
in the presence of an electric field with the help of the GAUSSIAN03 package.

\section{acknowledgements}
This work has been funded by Spanish MEC under Grants Nos. MAT2007-65487 and CSD2007-00010, and by
Generalitat Valenciana under Grant No. ACOMP07/054.
D.S. acknowledges financial support from Instituto de Cultura Juan Gil-Albert.
D.J. acknowledges financial support from the Spanish MEC under Grant No. UAC-2004-0052.

\begin{appendix}

\section{Derivation of Eq. (\ref{limit})}

In the limit of small lattice spacing $a$ the sums over the grid points $n$ in eq. (\ref{E-k-approx}) can be approximated by 
integrals over the quasi continuous variable $x=na$:
\begin{equation}
  \sum_{n=-\infty}^{\infty} t(na) \exp[ikna] \approx \frac{1}{a} \int{\rm d}x \, t(x) \, e^{ikx}
\end{equation}
and
\begin{equation}
  \sum_{n=-\infty}^{\infty} s(na) \exp[ikna] \approx \frac{1}{a} \int{\rm d}x \, s(x) \, e^{ikx}.
\end{equation}
Thus the two sums are approximately given by the Fourier transform of the hopping integral $t(x)$ and
the overlap integral $s(x)$, respectively. Since the overlap integral is a Gaussian function, 
$s(x)=\exp[-\alpha x^2]$, its Fourier transform $\tilde{s}(k)$ is also simply a Gaussian:
\begin{equation}
  \tilde{s}(k) = \frac{1}{\sqrt{2\pi}} \int{\rm d}x \, s(x) \, e^{ikx} = \alpha^{-1/2}\exp[-k^2/2\alpha].
\end{equation}

The Fourier transform $\tilde{t}(k)$ of the hopping integral $t(x)=\frac{\alpha}{2}(1-\alpha x^2)\exp[-\alpha x^2]$ 
is a bit more involved:
\begin{eqnarray}
  \tilde{t}(k) &=& \frac{1}{\sqrt{2\pi}} \int{\rm d}x \, t(x) \, e^{ikx} 
  \nonumber\\
  &=& \frac{\alpha}{2} \tilde{s}(k) - \frac{\alpha^2}{2}\frac{1}{\sqrt{2\pi}}\int{\rm d}x \, x^2 \, s(x) \, e^{ikx} 
  \nonumber\\
  &=& \frac{\sqrt{\alpha}}{2} \exp[-k^2/2\alpha] - i^2 \frac{\alpha^2}{2} \frac{{\rm d}^2}{{\rm d}k^2}\tilde{s}(k)
  \nonumber\\
  &=& \left( \frac{\sqrt{\alpha}}{2} + \frac{\alpha^2}{2} \alpha^{-1/2} \left( -\frac{1}{\alpha} + \frac{k^2}{\alpha^2} \right) \right) \exp[-k^2/2\alpha]
  \nonumber\\  
  &=& \frac{k^2}{2} \alpha^{-1/2} \exp[-k^2/2\alpha].
\end{eqnarray}

Thus, in total we obtain for the energy dispersion $\epsilon(k)$ in the limit of small $a$:
\begin{equation}
  \epsilon(k) = \frac{\sum_n t(na) \exp[ikna]}{\sum_n s(na) \exp[ikna]} \approx \frac{\tilde{t}(k)}{\tilde{s}(k)} = \frac{k^2}{2}.
\end{equation}
This proofs eq. (\ref{limit}) since in the limit $a\rightarrow0$ the approximation of the sums by integrals 
becomes exact.

\end{appendix}
      
\bibliography{matcon}

\begin{thebibliography}{26}
\bibitem[{mas()}]{mass:comment}
\bibinfo{note}{Alternatively, we could choose, e.g., the minimum of
  $m^\ast(k)$. However, this would result in only slightly lower values for
  $k_{\rm max}$ since the minimum occurs immediately before the singularity of
  $m^\ast(k)$ as can be seen from Fig.\ref{fig:mass}.}
\end{thebibliography}

\end{document}